\documentclass[aps,prl,twocolumn,bibnotes,groupedaddress]{revtex4}

\usepackage{amsmath,amsfonts,color,amssymb,mathrsfs,dsfont}

\newcommand{\dslash}{d \hspace{-0.8ex}\rule[1.2ex]{0.9ex}{.1ex}}
\newcommand{\deltaslash}{\delta \hspace{-0.8ex}\rule[1.2ex]{0.9ex}{.1ex}}

\begin{document}


\title{Duality and Decay of Macroscopic F-Strings}



\author{Dimitri P.~Skliros}

\author{Edmund J.~Copeland}

\author{Paul M.~Saffin}
\affiliation{School of Physics and Astronomy, University of Nottingham, \\Nottingham, Nottinghamshire NG7 2RD, UK}



\date{$10^{\rm th}$ June 2013}

\begin{abstract}
We study the decay of fundamental string loops of arbitrary size $L/{\rm min}(n,m)\gg\sqrt{\alpha'}$, labelled by $(n,m;\lambda_n,\bar{\lambda}_m)$, where $n,m$ correspond to left- and right-mover harmonics and $\lambda_n,\bar{\lambda}_m$ to polarisation tensors, and find that a description in terms of the recent coherent vertex operator construction of Hindmarsh and Skliros is computationally very efficient. We primarily show that the decay rates and mass shifts of vertex operators $(n,m;\lambda_n,\bar{\lambda}_m)$ and their ``duals'' $(n,m;\lambda_n,\bar{\lambda}_m^*)$ are equal to  leading order in the string coupling, implying for instance that decay rates of epicycloids equal those of hypocycloids. We then compute the power and decay rates associated to massless IR radiation for the trajectory $(1,1;\lambda_1,\bar{\lambda}_1)$, and find that it is precisely reproduced by the low energy effective theory of Dabholkar and Harvey. Guided by this correspondence, we conjecture the result for arbitrary trajectories $(n,m;\lambda_n,\bar{\lambda}_m)$ and discover a curious relation between gravitational and axion plus dilaton radiation. It is now possible to start exploring string evolution in regimes where a low energy effective description is less useful, such as in the vicinity of cusps.
\end{abstract}


\maketitle
There has been growing interest in the properties of macroscopic and/or excited string states, e.g.~in the context of AdS/CFT, see \cite{Tseytlin12} and references therein, where a precise understanding of the string/gauge theory spectrum is being explored, as a probe of black hole physics \cite{AmatiCiafaloniVeneziano}, in collider physics \cite{Feng:2010yx,Schlotterer11}  when the string scale is in the TeV range, and in higher-spin gauge theories \cite{SagnottiTaronna11}\cite{Schlotterer11}, where an intriguing  suggestion is that massive Regge excitations may arise from spontaneous breaking of a higher-spin gauge symmetry. We draw our main motivation from yet another arena, the phenomenologically interesting possibility of cosmic superstrings \cite{CopelandMyersPolchinski04,Hindmarsh11}, 
which provides a perfect context to probe stringy physics \cite{Hindmarsh11}. Recently, a description of generic excited strings in terms of covariant coherent vertex operators  became available \cite{HindmarshSkliros10,SklirosHindmarsh11}, and below we initiate a study of their interactions.

Total and partial decay widths of macroscopic and/or excited strings largely determine their phenomenological relevance at low energies. 
Most massive closed strings are unstable and decay \cite{WilkinsonTurokMitchell90,DabholkarMandalRamadevi98,IengoRusso02,ChialvaIengoRusso03,GutperleKrym06,IengoRusso06} \cite{Manes02,ChialvaIengoRusso05}, typical decay rates being (when massive radiation is negligible), 
$$
\Gamma \sim G_D\mu^2(M/\mu)^{5-D},
$$ 
with $G_D$ Newton's constant in $D$ non-compact dimensions, $\mu=1/(2\pi\alpha')$ the string tension and $M$ the mass of the decaying string. The associated lifetime is, 
$$
\tau\sim (G_D\mu)^{-1}(M/\mu)^{D-3},
$$
and so low GUT scale strings the size of a galaxy, the solar system, or an atom would survive $\tau\sim10^{14}\,$y,  $10^4\,$y, and $10\,$ps respectively in $D=4$. There are also more stable near-BPS trajectories (see e.g.~\cite{GutperleKrym06}) and approximate non-renormalisation theorems \cite{DabholkarMandalRamadevi98}. Stable BPS states in turn provide a window into non-perturbative physics \cite{Sen98b}\cite{classicaleffective,DasMathur96b,BianchiLopezRichter10}. D-string decay at strong coupling can also be extracted from F-string decay at weak coupling \cite{DasMathur96b}, whereas a handle at finite coupling may also be within reach \cite{Sen13}. 

Once formed, unstable strings provide a source of radiation that generically is strongly anisotropic \cite{DamourVilenkin00,VachaspatiVilenkin85,Burden85}. Given a large enough sample of such strings, a classical analysis (which however neglects radiative backreaction \cite{QuashnockSpergel90} and massive emission) shows that we may be able to observe this radiation in the form of gravitational wave bursts \cite{DamourVilenkin00}, or cosmic rays, see e.g.~\cite{Hindmarsh11}, and so predicting the precise spectrum is crucial. 

Below we study the decay of highly excited fundamental string loops.  Traditionally, these are described in a mass eigenstate basis, and calculations almost exclusively \cite{WilkinsonTurokMitchell90,DabholkarMandalRamadevi98,IengoRusso02,Manes02,ChialvaIengoRusso03,ChialvaIengoRusso05,GutperleKrym06,IengoRusso06} focus on leading Regge trajectory states which do not exhibit generic features (such as non-degenerate cusps \cite{HindmarshKibble95}). Although physical vertex operators for arbitrary \emph{mass eigenstates} are known \cite{SklirosHindmarsh11}, they are not a natural basis for highly excited strings as the resulting amplitudes generically become unwieldy. Below we show that a  \emph{coherent state} basis \cite{HindmarshSkliros10}, a complete set of which was constructed in \cite{SklirosHindmarsh11}, is much more natural. 
For worldsheet embeddings $X^M:\Sigma\rightarrow \mathbb{R}^{D-1,1}\times T^{26-D}$, in a flat string background, $G_{MN}=\eta_{MN}$, with vanishing Kalb-Ramond 2-form $B_{(2)}=0$ and constant dilaton $\Phi= \Phi_0$ (with string coupling $e^{\Phi_0}\ll1$), we consider the coherent vertex operators \cite{HindmarshSkliros10,SklirosHindmarsh11}: 
\begin{equation}\label{eq:(n,m) coherent vertex}
\begin{aligned}
\mathcal{O}(z,&\bar{z})= \,\,:\!C\int_0^{2\pi} \!\!\dslash s \exp\Big(\frac{i}{n}\,e^{ins}\lambda_n\cdot D^{n}_zX\,e^{-inq\cdot X(z)}\Big)\\
&\times \exp\Big(\frac{i}{m}\,e^{-ims}\bar{\lambda}_m\cdot \bar{D}^{m}_{\bar{z}}X\,e^{-imq\cdot X(\bar{z})}\Big)\,e^{ip\cdot X(z,\bar{z})}\!:
\end{aligned}
\end{equation} 
when $\lambda_n\cdot p=0$ and $p^{\mu},q^{\mu}\in\mathbb{R}^{D-1,1}$. The polarisation tensors, $\lambda_n^i$, span a $\mathbb{R}^{D-2}$ subspace. 
$\mathcal{O}(z,\bar{z})$ will be a $(1,1)$ conformal primary \cite{Witten12c} when \cite{SklirosHindmarsh11} $\lambda_{n}^2=\lambda_n\cdot q=0$, $p^2=2p\cdot q=4/\alpha'$, and $q^2=0$.  (Identical remarks hold for the anti-chiral half, $\bar{\lambda}_m^i$.) 

We define 
$
D^{n}_z\equiv  \sum_{r=1}^n\frac{S_{n-r}(a_s)}{(r-1)!}\,\partial_z^r,
$ 
with $S_m(a_s)$ elementary Schur polynomials \footnote{Elementary Shur polynomials, $S_m(a_s)$, are defined by the generating function $\sum_{m=0}^{\infty}S_m(a_s)u^m\equiv \exp \sum_{s=1}^{\infty}a_s\,u^s$.} and $a_s= -\frac{1}{s!}inq\cdot\partial^sX(z)\label{eq:S_m(a)b}$, and likewise for the anti-holomorphic quantities \cite{SklirosHindmarsh11}. The operator-product expansion (OPE), $\mathcal{O}^{\dagger}(z,\bar{z})\cdot \mathcal{O}(0,0)\cong g_D^2/|z|^4$, fixes the normalisation \footnote{Here, $C^{-2} = g_D^{-2}\int_0^{2\pi} \frac{ds}{2\pi}\,\exp\big(e^{ins}\frac{|\lambda_n|^2}{n}\big)\exp\big(e^{-ims}\frac{|\bar{\lambda}_m|^2}{m}\big)$. In terms of the compactification volume, $g_{D}\equiv \frac{{\alpha'}^6e^{\Phi_0}}{\sqrt{V_{26-D}}}$.} and unitarity \cite{Polchinski_v1} sets $\big(2\pi g_D)^2=8\pi G_{D}$.

The momentum expectation value of $\mathcal{O}(z,\bar{z})$ is $\langle \hat{p}^{\mu}\rangle=p^{\mu}-\ell^2 q^{\mu}$, with $\ell\equiv |\lambda_n|=|\bar{\lambda}_m|$. We work throughout in the ``rest frame'' \cite{SklirosHindmarsh11}, $\langle \hat{p}^{N}\rangle= M\,\delta^{N}_{\phantom{a}0}$, with $M\equiv \mu L$ the invariant mass. The string length, $L$, is then determined from the Virasoro constraints, $L=4\pi\sqrt{\alpha'\ell^2-\alpha'}$. 

When $\ell\gg1$, rest-frame vertex operators (\ref{eq:(n,m) coherent vertex}) are in one-to-one correspondence with classical string trajectories \cite{HindmarshSkliros10}, 
$X^{N}=(X^0,X^i,X^{D-1},X^a)$, where $X^0= -iM\ln z\bar{z}$,
\begin{equation}\label{eq:<X>_coherent_lc closed}
\begin{aligned}
&X^i=\frac{i}{n}\,\big(\lambda_n^i\,z^{-n}\!\!-\lambda_n^{*i}\,z^{n}\big)+\frac{i}{m}\big(\bar{\lambda}_m^i\,\bar{z}^{-m}\!\!-\bar{\lambda}_m^{*i}\,\bar{z}^{m}\big),
\end{aligned}
\end{equation}
and $X^a=X^{D-1}= 0$, which are closely related to the trajectories in \cite{Burden85}; here $\alpha'=2$. The indices $\mu=(0,i,D-1)$ and $a$ span $\mathbb{R}^{D-1,1}$ and $T^{26-D}$ respectively. When the polarisation tensors satisfy the aforementioned relations, the equations of motion, $\partial_z \partial_{\bar{z}}X^{N}=0$, and constraints, $(\partial_z X)^2=(\partial_{\bar{z}}X)^2=0$, are satisfied. For general harmonics, $n,m$, with greatest common divisor $g$, denoted by $[n,m]=g$, there exist unique relatively-prime positive integers, $u,w$, such that $n=gu$ and $m=gw$, with $[u,w]=1$. Redefining $z,\bar{z}\rightarrow z^g,\bar{z}^g$, we learn that loops distinguished by different values of $g$ (with $u,w$ fixed) are self-similar: increasing the ``winding'' number \footnote{Notice that (\ref{eq:<X>_coherent_lc closed}) is invariant under $g\rightarrow g+1$, up to a rescaling of $L$ and a worldsheet reparametrisation. Furthermore, $M\sim g\mathcal{R}/\alpha'$, which is characteristic of winding modes; here $\mathcal{R}$ is determined by the string dynamics.} $g$ decreases their size, $\sim L/g$, their shape remaining fixed. 

Given a set of trajectories $X=X_{\rm L}(z)+X_{\rm R}(\bar{z})$, the distinct trajectories $X'=X_{\rm L}(z)-X_{\rm R}(\bar{z}^{-1})$ are also physical; we call the latter \emph{dual} trajectories, obtained from (\ref{eq:<X>_coherent_lc closed}) or (\ref{eq:(n,m) coherent vertex}) by $(n,m;\lambda_n,\bar{\lambda}_m)\rightarrow  (n,m;\lambda_n,\bar{\lambda}_m^*)$, see also \cite{MosaffaSafarzadeh07IshizekiKruczenski07,BerkovitzMaldacena08KruczenskiTseytlin12}. 

To visualise some of the trajectories captured by, $\mathcal{O}(z,\bar{z})$, consider $D=4$ and pick a coordinate system such that $\lambda_n = \frac{1}{\sqrt{2}}\,\ell(\hat{\bf x}+i\hat{\bf y})$ and $\bar{\lambda}_m =\frac{1}{\sqrt{2}}\,\ell(\!-\hat{\bf x}+ i\cos\psi \,\hat{\bf y})$, with $\psi  = 0$ or $\pi$.  The angular momentum is \cite{SklirosHindmarsh11}:
$$
\langle J^{xy}\rangle= \frac{1}{2}\Big(\frac{1}{n}-\frac{\cos\psi}{m}\Big)\Big(\frac{\alpha'}{2}\,M^2+2\Big),
$$ 
giving rise to leading Regge and various ``sister'' trajectories.  When $\ell\gg1$, the {\it rms} string radius is \cite{SklirosHindmarsh11}: 
$$
\mathcal{R}=\frac{L}{4\pi\sqrt{2}}\sqrt{\frac{1}{n^2}+\frac{1}{m^2}-\frac{\delta_{n,m}}{nm}(1+\cos\psi)\cos \frac{4\pi nX^0}{L}}.
$$
Trajectories self-intersect if both $u,w>1$ \cite{Burden85}. We label solutions by $(n,m;\psi)$, with two subclasses: $(n,m;0)$ and $(n,m;\pi)$. 
If $m$ is a divisor of $n$ (i.e.~$g=m$), $(n,m;0)$ are \emph{epicycloids} (with $n/m-1$ cusps \footnote{A cusp is a point on $\Sigma$ at which the determinant of the embedding metric vanishes, $\partial_zX\cdot \partial_{\bar{z}}X=0$.}) and $(n,m;\pi)$ \emph{hypocycloids} (with $n/m+1$ cusps), the two being dual in the above sense \footnote{This was referred to as T-duality in \cite{MosaffaSafarzadeh07IshizekiKruczenski07}, although this identification is subtle  in non-compact spacetime \cite{BerkovitzMaldacena08KruczenskiTseytlin12} and a general Lorentz frame.}. The simplest example of the former is a pulsating circle, $(1,1;0)$, followed by a rotating cardioid, $(2,1;0)$, etc., whereas the simplest example of the latter is a rotating folded string, $(1,1;\pi)$, followed by a rotating deltoid, $(2,1;\pi)$, etc., and similarly for $g=n$. In $D=4$, all trajectories except $(n,n;0)$ have permanent cusps; in $D>4$, if $\lambda_n\cdot \bar{\lambda}_m^*$ and $\lambda_n\cdot \bar{\lambda}_m$ are \emph{either} real or imaginary, cusps generically appear at discrete instants \cite{CopelandSaffinSkliros13}.

Decay rates may be extracted from the imaginary part of two-point amplitudes  \cite{WilkinsonTurokMitchell90,DabholkarMandalRamadevi98,IengoRusso02,ChialvaIengoRusso03,GutperleKrym06,IengoRusso06}, $\Gamma = \frac{1}{M}\,{\rm Im}\,\mathcal{M}$. The amplitudes $\mathcal{M}$ are in turn obtained by integrating over complex moduli, $\tau$, and vertex insertion points, $z$, of a compact Riemann surface, and over matter, $X$, and $b,c$ ghost contributions \cite{DHokerPhong,VerlindeVerlinde_lec,Witten12c}. When strings in the loops go onshell, they can appear in final states. In the usual approach \cite{WilkinsonTurokMitchell90,DabholkarMandalRamadevi98,IengoRusso02,ChialvaIengoRusso03,GutperleKrym06,IengoRusso06}, dependence on the loop momenta is not manifest, and so one does not have a handle on the frequency and isotropy of the emitted radiation. We bypass this obstacle by considering instead the fixed ($A$-cycle) loop momentum amplitude \cite{DHokerPhong}, $\mathcal{M}(\mathbb{P})$, defined (at genus-1) by $\mathcal{M}\equiv \int \frac{d^D\mathbb{P}}{(2\pi)^D}\mathcal{M}(\mathbb{P})$, in terms of which, 
\begin{equation}\label{eq:dP/dO}
\begin{aligned}
\Gamma = \frac{1}{\mu L}\int \frac{d^D \mathbb{P}}{(2\pi)^D}\,\textrm{Im}\,\mathcal{M}(\mathbb{P}).
\end{aligned}
\end{equation} 
The total power emitted, $P$, is in turn obtained by means of $dP=\mathbb{P}^0d\Gamma$. The quantities of interest will be the decay rates, $d\Gamma/d\Omega_{S^{D-2}}$, and power emitted, $dP/d\Omega_{S^{D-2}}$, per unit solid angle. We extract the contribution associated to decay channels of interest by factorizing $\mathcal{M}(\mathbb{P})$ on the corresponding poles, e.g.~$\mathbb{P}^2=0$ for massless radiation. 

It is worth emphasising that the fixed-loop momenta amplitudes, $\mathcal{M}(\mathbb{P})$, are also more natural than their integrated manifestations, $\mathcal{M}$, in light of a chiral splitting theorem \cite{DijkgraafVerlindeVerlinde88}, and this will play an important role below.

To determine $\mathcal{M}(\mathbb{P})$, note that the ghost contribution amounts to an overall factor $|\eta(\tau)|^{4}$, with $\eta(\tau)$ the Dedekind eta function \cite{DHokerPhong}. We then extract $\mathcal{M}(\mathbb{P})$ by inserting a momentum-conserving delta function into the path integral which only allows strings of momentum $\hat{\mathbb{P}}=\mu\oint_{A} (\partial X-\bar{\partial}X)$ to propagate through the loop \footnote{We use a canonical intersection basis \cite{DHokerPhong}, defined such that $\oint_Adz=1$, $\oint_Bdz=\tau$, with identifications $z\sim z+1$, $z\sim z+\tau$ and $\tau$ the modulus of the torus},
\begin{equation}\label{eq:A1}
\begin{aligned}
i\deltaslash^D(0&)\mathcal{M}(\mathbb{P})= \frac{1}{2}\int_{\mathcal{M}_1} \!\!d^2\tau|\eta(\tau)|^4\int_{\Sigma}d^2z\\
&\times\int \mathcal{D}X\,e^{-S_X}\deltaslash^{D}\big(\mathbb{P}-\hat{\mathbb{P}}\big)\mathcal{O}^{\dagger}(z,\bar{z})\mathcal{O}(z',\bar{z}'),
\end{aligned}
\end{equation}
with $z',\bar{z}'$ fixed. Our gauge slice choice leads to a moduli space \footnote{Our measure conventions are $d^2z\equiv idz\wedge d\bar{z}$, $d^2\tau\equiv id\tau\wedge d\bar{\tau}$, $\int \mathcal{D}(\delta X)e^{-\frac{1}{2}\mu\int d^2z\sqrt{g}(\delta X)^2}=1$ and $d \hspace{-0.3ex}\bar{}\,s\equiv ds/(2\pi)$, $\delta\hspace{-0.3ex}\bar{}\,(x)=(2\pi)\delta(x)$.\label{ft:measure}} $\mathcal{M}_1\equiv \{|\tau|^2\geq1,\,-\frac{1}{2}\leq {\rm Re}\,\tau\leq \frac{1}{2},\,{\rm Im}\,\tau>0\}$ and $\Sigma=\mathbb{C}/(\mathbb{Z}+\tau\mathbb{Z})$, with metric $ds^2=dzd\bar{z}$. The remaining $\mathbb{Z}_2$ isometry leads to the factor of $1/2$ in (\ref{eq:A1}), although the overall normalisation is fixed by unitarity \cite{Polchinski88}. The Euclidean string action is, 
$
S_X=\mu \int_{\Sigma}d^2z\partial_z X^{\mu}\partial_{\bar{z}}X^{\nu}\delta_{\mu\nu},
$ 
and Hermitian conjugation takes $(p^{\mu},q^{\mu}; \lambda_n,\bar{\lambda}_m)\rightarrow(-p^{\mu},-q^{\mu};-\lambda_n^*,-\bar{\lambda}_m^*)$ in (\ref{eq:(n,m) coherent vertex}). 

We use point splitting to integrate out $X$ \cite{KohTroostVanProeyen87}. Paying careful attention to the combinatorics and normalisation, the chirally-split \cite{DijkgraafVerlindeVerlinde88} two-point coherent state amplitude reads \cite{SklirosCopelandSaffin13}:
\begin{equation}\label{eq:M(P)full nm}
\begin{aligned}
\mathcal{M}(\mathbb{P})= \sum_{K,W}\frac{1}{2}\int_{\mathcal{M}_1}d^2\tau\int_{\Sigma}d^2z\int_0^{2\pi}\!\! \dslash s\,\mathcal{F}_n(z|\tau)\,\bar{\mathcal{F}}_m(\bar{z}|\bar{\tau}),
\end{aligned}
\end{equation}
where, writing Fay's prime form \cite{DHokerPhong} as $E=E(z,z')$, the chiral half reads,
\begin{equation}\label{eq:Fn(z|t)}
\begin{aligned}
&\mathcal{F}_n(z|\tau)\equiv\,\, C\,\eta(\tau)^{-24}e^{\pi i\tau \mathbb{P}^2}E^{-p^2}e^{-2\pi i\,\mathbb{P}\cdot p(z-z')}\\
&\times\exp\bigg\{e^{ins}\frac{1}{n^2}|\lambda_n|^2\,e^{2\pi i\mathbb{P}\cdot nq(z-z')}E^{2n}\mathcal{D}_z^n\mathcal{D}_{z'}^n\ln E\bigg\}\\
&\times I_0\Big(2\,e^{\frac{ins}{2}}\frac{1}{n}\,|\mathbb{P}\cdot \lambda_n|\,e^{\pi i\mathbb{P}\cdot nq(z-z')}2\pi E^{n}\,\mathcal{S}_{n-1}\Big).
\end{aligned}
\end{equation}
The quantities $\mathcal{S}_{n-\ell}$, $\mathcal{D}_z^n$ are equal to $S_{n-\ell}$, $D_z^n$ above, but now $a_s\equiv \frac{n}{s!}\,\partial_z^s\big(\ln |E|^2-4\pi\mathbb{P}\cdot q{\rm Im}\,(z-z')\big)$. The $I_0(2x)$ are modified Bessel functions. The integers $K,W\in\mathbb{Z}^{26-D}$ are momentum and winding modes of $A$-cycle strings associated to $T^{26-D}$. The momenta are $\mathbb{P}^a\equiv \big(\frac{K}{R}\big)^a+\frac{(WR)^a}{2}$ and $\bar{\mathbb{P}}^a\equiv \big(\frac{K}{R}\big)^a-\frac{(WR)^a}{2}$, where $x^a\sim x^a+2\pi R^a$. The anti-chiral half, $\bar{\mathcal{F}}_m(\bar{z}|\bar{\tau})$, is in turn obtained by complex conjugation from $\mathcal{F}_n(z|\tau)$ and the replacements $(n,\lambda_n,\mathbb{P})\rightarrow (m,\bar{\lambda}_m,\bar{\mathbb{P}})$ \footnote{The momenta are to be assumed real, but may subsequently be analytically continued to the complex plane.}.

The normalisation of (\ref{eq:M(P)full nm}) has been fixed by unitarity \cite{SklirosCopelandSaffin13}: we factorize ${\rm Im}\,\mathcal{M}(\mathbb{P})$ on two tachyon poles, take $\ell\rightarrow0$, and relate \cite{Polchinski88} this to the modulus squared of the three-tachyon amplitude \cite{Polchinski_v1}, $\mathcal{M}_{ttt}=8\pi g_D/\alpha'$, using the relation, $
2\,{\rm Im}\,\mathcal{M}(\mathbb{P})=\frac{1}{2}\sum_{m^2,M^2,{\rm pol.}}\,\deltaslash(\mathbb{P}^2+m^2)\deltaslash\big((k-\mathbb{P})^2+M^2\big)\theta(\mathbb{P}^0)\theta(k^0-\mathbb{P}^0)|\mathcal{M}^{(\rm tree)}|^2
$. (The symmetry factor of $1/2$ is only present for  indistinguishable particles in the loop.)

Notice that (\ref{eq:M(P)full nm}) is invariant under $(n,m,\lambda_n,\bar{\lambda}_m)\rightarrow(n,m,\lambda_n,\bar{\lambda}_m^*)$, a powerful (albeit tree-level) statement relating decay rates (and mass shifts) of \emph{distinct} states,
\begin{equation}\label{eq:duality}
\frac{d\Gamma}{d\Omega_{S^{D-2}}}\Big|_{(n,m,\lambda_n,\bar{\lambda}_m)}=\frac{d\Gamma}{d\Omega_{S^{D-2}}}\Big|_{(n,m,\lambda_n,\bar{\lambda}_m^*)},
\end{equation}
so for instance, in $D=4$, decay rates of epicycloids $(gu,g;0)$ and the dual hypocycloids $(gu,g;\pi)$, remarkably, are equal, for all $g,u\in\mathbb{Z}^+$. 
This also generalises the observation of Iengo and Russo  \cite{IengoRusso06}, that decay rates of pulsating circles and folded strings are equal, and remains true (at one loop) for all vertex operators, $\mathcal{O}(z,\bar{z})$. 

As an example, let us compute (\ref{eq:dP/dO}) explicitly for massless radiation for the trajectory $(1,1,\lambda_1,\bar{\lambda}_1)$ and its dual $(1,1,\lambda_1,\bar{\lambda}_1^*)$. The integrand of (\ref{eq:M(P)full nm}) consists of a linear superposition of terms  
$|e^{2\pi i(\tau-z)}|^{\mathbb{P}^2+m_1^2}|e^{2\pi iz}|^{(k-\mathbb{P})^2+{m_2}^2}$, 
there being a sum over onshell masses, $m_1^2,m_2^2$, of states that propagate in the loop, and over mass eigenstates of momentum $k=p-aq$, with $a=0,1,2,\dots$, of which the decaying coherent state is composed. We search for branch cuts in the complex $k^0$ plane, the discontinuity across which yields an imaginary part, ${\rm Im}\mathcal{M}(\mathbb{P})$ \cite{LandauLifshitzRQF}, whereby we analytically continue back to Minkowski signature. Only the $m_1^2=0$ and ${m_2}^2=0$ terms contribute to massless radiation. We write the associated onshell $D$-momentum of the radiation as $\mathbb{P}^{\mu}=\mathbb{P}^0(1,\hat{\mathbb{P}}^j)$ with $\hat{\mathbb{P}}^2=1$. Omitting the details of the computation \cite{SklirosCopelandSaffin13}, in the IR region of the emission spectrum, $\mathbb{P}^0\ll\mu L$, the resulting series can be resummed , leading precisely to the following expression for the power per unit solid angle in a direction $\hat{\mathbb{P}}$ into massless radiation of energy $\mathbb{P}^0=4\pi Nuwg/L$, with $N=1,2\dots$: 
\vspace{-1cm}
\begin{widetext}
\begin{equation}\label{eq:dP/dOaprxQ}
\begin{aligned}
\frac{dP_N}{d\Omega_{S^{D-2}}}\Big|_{m^2=0}\!\!\!\!
&=\frac{16\pi G_D\mu^2}{(2\pi)^{D-4}}\Big(\frac{4\pi Nuwg}{L}\Big)^{D-4-\delta}\!\!\!(Nuwg)^2\Big[{J'}^2_{Nw}\big(Nw\sqrt{2}|\hat{\mathbb{P}}\cdot\hat{\lambda}_n|\big)+\Big(\frac{1}{2}|\hat{\mathbb{P}}\cdot\hat{\lambda}_n|^{-2}-1\Big)J_{Nw}^2\big(Nw\sqrt{2}|\hat{\mathbb{P}}\cdot\hat{\lambda}_n|\big)\Big]\\
&\hspace{1.3cm}\quad\times\Big[{J'}^2_{Nu}\big(Nu\sqrt{2}|\hat{\mathbb{P}}\cdot\hat{\bar{\lambda}}_m|\big)+\Big(\frac{1}{2}|\hat{\mathbb{P}}\cdot\hat{\bar{\lambda}}_m|^{-2}-1\Big)J_{Nu}^2\big(Nu\sqrt{2}|\hat{\mathbb{P}}\cdot\hat{\bar{\lambda}}_m|\big)\Big],\\
\end{aligned}
\end{equation}
\end{widetext}
\vspace{-1cm}
with $u,w,g=1$, $\lambda_n=\ell\hat{\lambda}_n$, $\bar{\lambda}_m=\ell\hat{\bar{\lambda}}_m$, and $\delta=0$. When $\delta=1$ the above yields a decay rate ($d\Gamma_N/d\Omega_{S^{D-2}}$, probability per unit time per unit solid angle that a massless particle of energy $\mathbb{P}^0$ is emitted in direction $\hat{\mathbb{P}}$). We conjecture that (\ref{eq:dP/dOaprxQ}) hold for arbitrary non self-intersecting trajectories $(n,m;\lambda_n,\bar{\lambda}_m)$ provided perturbation theory remains valid. (The integers $u,w,g$ are defined in terms of $n,m$ above and the $J_n(z)$ are Bessel functions.)

The $L$ dependence, $\Gamma\sim G_D\mu^2L^{5-D}g^{D-3}$, agrees with \cite{IengoRusso02} in $D=10$. When massless radiation is the dominant decay channel, the $L$ dependence of the total lifetime, $\tau\sim (G_D\mu g)^{-1}(L/g)^{D-3}$, is in agreement with \cite{ChialvaIengoRusso05} in $D=10$ and \cite{VilenkinShellard94} in $D=4$, but see also \cite{WilkinsonTurokMitchell90}. 

The characteristic cusp signal \cite{VachaspatiVilenkin85,DamourVilenkin00} comes from the region $\sqrt{2}|\hat{\mathbb{P}}\cdot\hat{\lambda}_n|=\sqrt{2}|\hat{\mathbb{P}}\cdot\hat{\bar{\lambda}}_m|=1$, which for large $N$ leads to 
$$
\frac{dP_N}{d\Omega_{S^{D-2}}}\propto (Nuw)^{D-4-\delta}\Big(\frac{uw}{N}\Big)^{2/3}g^{D-2-\delta},
$$ 
generalising the $D=4$ result  \cite{VachaspatiVilenkin85}, $dP_N/d\Omega_{S^2}\propto N^{-2/3}$, see also \cite{Iengo06}. For $D=4$, the explicit polarisation tensors are given below (\ref{eq:<X>_coherent_lc closed}), and we write $\hat{\mathbb{P}}^i=\sin\theta \cos\varphi \hat{\bf x}+\sin\theta\sin\varphi\hat{\bf y}+\cos\theta\hat{\bf z}$ in (\ref{eq:dP/dOaprxQ}), from which $\sqrt{2}|\hat{\mathbb{P}}\cdot\hat{\lambda}_n|=\sqrt{2}|\hat{\mathbb{P}}\cdot\hat{\bar{\lambda}}_m|=\sin\theta$, with $\theta\in[0,\pi]$ -- the radiation is axially symmetric about the $z$ axis.
\vspace{-0cm}

Remarkably, (\ref{eq:dP/dOaprxQ}) (with $\delta=0$) is precisely equal \cite{CopelandSaffinSkliros13} to the power per unit solid angle associated to $G+B+\Phi$ emission, as derived from the following effective action in the absence of radiative backreaction \cite{classicaleffective,CopelandHawsHindmarsh90}:
\vspace{-0.6cm}
\begin{widetext}
\begin{equation}\label{eq:Seff}
\begin{aligned}
&S_{\rm eff} =\frac{1}{16\pi G_{D}}\int d^Dx\sqrt{-G}\,e^{-2\Phi}\Big(R_{(D)}+4(\nabla\Phi)^2-\frac{1}{12}\,H_{(3)}^2+\dots\Big)-\mu\int_{S^2} \!\partial X^{\mu}\wedge\bar{\partial} X^{\nu}\big(G_{\mu\nu}+B_{\mu\nu}\big)+\dots
\end{aligned}
\end{equation}
\end{widetext}
In particular, we perturb around the string background, $G_{\mu\nu}\simeq \eta_{\mu\nu}+h_{\mu\nu}$, $B_{\mu\nu}\simeq b_{\mu\nu}$ and $\Phi\simeq \Phi_0+\phi$, in (\ref{eq:Seff}) and evaluate the source term on the trajectories (\ref{eq:<X>_coherent_lc closed}). The power is extracted from the energy-momentum, $T^{\mu\nu}_{G+B+\Phi}$, carried away from the string source by the perturbations $h_{\mu\nu},b_{\mu\nu}$ and $\phi$, according to \cite{WeinbergGC,Burden85} 
$
dP/d\Omega_{S^{D-2}} = \big\langle r^{D-2}\hat{\mathbb{P}}^i T^{0i}_{G+B+\Phi}\big\rangle
$, with $r$ the distance to the observer. The averaging is over spacetime regions much larger than the wavelengths under consideration \cite{WeinbergGC}, and the resulting expression is precisely (\ref{eq:dP/dOaprxQ}) (in \emph{both} the string and Einstein frame \footnote{There is an interesting story here: in the string frame dilaton radiation vanishes, whereas the gravitational radiation is correspondingly larger \cite{CopelandSaffinSkliros13}.}, the two being related by $G_{\mu\nu}=e^{4\phi/(D-2)}G_{\mu\nu}^{\rm E}$.)

The power into gravitational waves, in $D=4$, was computed in \cite{Burden85}, and for $D\geq4$ in \cite{CopelandSaffinSkliros13}. 
Let us write $\mathcal{O}=(n,m,\lambda_n,\bar{\lambda}_m)$, $\tilde{\mathcal{O}}=(n,m,\lambda_n,\bar{\lambda}_m^*)$. 
Intriguingly, the power into massless radiation equals the sum of the gravitational wave results \cite{Burden85} for the dual trajectories, 
\begin{equation}\label{eq:Q=1/2(C1+C2)}
\begin{aligned}
\frac{dP}{d\Omega}\Big|^{\mathcal{O}}_{m^2=0}&=\frac{dP}{d\Omega}\Big|^{\tilde{\mathcal{O}}}_{m^2=0} = \frac{dP}{d\Omega}\Big|_{G}^{\mathcal{O}}+\frac{dP}{d\Omega}\Big|_{G}^{\tilde{\mathcal{O}}},
\end{aligned}
\end{equation}
where we note that 
$$
\frac{dP}{d\Omega}\Big|_{m^2=0}=\frac{dP}{d\Omega}\Big|_{G}+\frac{dP}{d\Omega}\Big|_{B}+\frac{dP}{d\Omega}\Big|_{\Phi}.
$$ 

To conclude, we have constructed the two-point amplitudes (\ref{eq:M(P)full nm}), associated to coherent vertex operators (\ref{eq:(n,m) coherent vertex}), labelled by quantum numbers $(n,m,\lambda_n,\bar{\lambda}_m)$, in $D$ non-compact dimensions. There is an invariance under $(n,m,\lambda_n,\bar{\lambda}_m)\rightarrow (n,m,\lambda_n,\bar{\lambda}_m^*)$, implying a symmetry (\ref{eq:duality}) (and (\ref{eq:Q=1/2(C1+C2)})). This is possibly broken by higher loop effects: the quantity $(\mathbb{P}_I\cdot \lambda_n^*)\mathcal{D}_z^n\int^z\!\!\omega_I \,(\mathbb{P}_J\cdot \lambda_n)\mathcal{D}_{z'}^n\int^{z'}\!\!\omega_J$ that would appear in the argument of the Bessel function in (\ref{eq:Fn(z|t)}) when $h\geq 1$, with $\omega_I$ ($I=1,\dots,h$) the holomorphic abelian differentials of the genus-$h$ Riemann surface and $\mathbb{P}^{\mu}_I$ the momentum through the $A_I^{\rm th}$ homology cycle \cite{DHokerPhong}, is only invariant when $h=1$.

We have derived the power and decay rate (\ref{eq:dP/dOaprxQ}) associated to massless IR radiation for the trajectories $(1,1;\lambda_1,\bar{\lambda}_1)$, when backreaction is neglected, and find that the result can be precisely reproduced by a low energy effective theory \cite{classicaleffective}. We conjecture the result for general (non self-intersecting) trajectories $(n,m;\lambda_n,\bar{\lambda}_m)$, provided perturbation theory remains valid \footnote{In $D=4$, string loops $(n,m;\lambda_n,\bar{\lambda}_m)$ will remain larger than their Schwarzchild radius if ${\rm min}(n,m)\ll (8\pi G_4\mu)^{-1}$; for low GUT scale strings, ${\rm min}(n,m)\ll 10^6$.}. 

The coherent state basis \cite{HindmarshSkliros10} and fixed-loop momenta approach is seemingly very efficient for string calculations, and makes it possible to start exploring stringy effects (such as cusp emission \cite{DamourVilenkin00,Iengo06} and the effects of massive string modes) in regimes where the low energy effective description is less usefull. Due to the universality of (\ref{eq:Seff}) \cite{Polchinski_v1}, we expect that (\ref{eq:dP/dOaprxQ}) holds also for the superstring. 

\textit{Acknowledgements}: DPS thanks Mark Hindmarsh who contributed significantly to the seeds of this project and for various comments on the manuscript, and Eric D'Hoker, Ue-Li Pen, Jorge Russo, and especially Joe Polchinski and Arkady Tseytlin for fruitful discussions and suggestions. EJC acknowledges support from the Royal Society, Leverhulme Trust and, along with PMS and DPS, from STFC. DPS also acknowledges support from the University of Nottingham.

\bibliography{spi}

\end{document}